\documentclass{naturesmall}
\usepackage{fullpage,txfonts,hyperref}
\usepackage[stable]{footmisc}
\usepackage{graphicx}
\usepackage{color}

\newcommand{\lya}{Ly$\alpha$}
\newcommand{\hi}{H~{\sc i}}

\newcommand{\gt}{$>$}
\newcommand{\lt}{$<$}

\newcommand\aj{Astron. J.}%
\newcommand\apj{Astrophys J.}%
\newcommand\apjl{Astrophys J. Lett.}%
\newcommand\aap{Astron. Astrophys.}%
\newcommand\mnras{Mon. Not. R. Astron. Soc.}%

\title{Central Powering of the Largest Lyman-alpha Nebula is Revealed by Polarized Radiation}
\author{Matthew Hayes$^{1,2,3}$, Claudia Scarlata$^{4,5}$, and Brian Siana$^{6}$}

\begin{document}
\maketitle

\begin{affiliations}
\item Universit\'e de Toulouse, UPS-OMP; IRAP; Toulouse, France
\item CNRS; IRAP; 14, avenue Edouard Belin, F-31400 Toulouse, France
\item Observatory of Geneva, University of Geneva, 51 ch. des Maillettes, 1290 Versoix, Switzerland
\item Minnesota Institute for Astrophysics, School of Physics and Astronomy, University of Minnesota, Minneapolis, MN 55455, USA 
\item Spitzer Science Center, California Institute of Technology, 220-6, Pasadena, CA 91125, USA
\item Department of Astronomy, California Institute of Technology, MS 249-17, Pasadena, CA 91125, USA 
\end{affiliations}

\begin{abstract}
\textbf{High-redshift Lyman-alpha blobs\cite{Francis1996,Steidel2000} are extended, luminous,
but rare structures that appear to be associated with the highest peaks in the 
matter density of the Universe\cite{Palunas2004,Matsuda2004,Yang2009,Prescott2008}.
Their energy output and morphology are similar to powerful radio galaxies\cite{Saito2006}, but 
the source of the luminosity is unclear. Some blobs are 
associated with ultraviolet or infrared bright galaxies, suggesting an extreme
starburst event or accretion onto a central black 
hole\cite{Geach2009,Scarlata2009,Colbert2011}.
Another possibility is gas that is shock excited by 
supernovae\cite{Taniguchi2000,Mori2004}. 
However some blobs are not associated with 
galaxies\cite{Nilsson2006,SmithJarvis2007}, 
and may instead be heated by gas falling into a dark matter 
halo\cite{Fardal2001,Yang2006,DijkstraLoeb2009,Goerdt2010,Faucher2010}. 
The polarization of the \lya\ emission can in principle distinguish between 
these options\cite{LeeAhn1998,DijkstraLoeb2008,RybickiLoeb1999}, 
but a previous attempt to detect this signature 
returned a null detection\cite{Prescott2011}. 
Here we report on the detection of polarized \lya\ from the blob 
LAB1\cite{Steidel2000}. Although the central region shows no measurable polarization,
the polarized fraction 
 ($P$) increases to $\approx 20$ per cent at a radius of 45~kpc, 
forming an almost complete polarized ring. 
The detection of polarized radiation
is inconsistent with the in situ production of \lya\ photons, and we conclude 
that they must have been produced in the galaxies hosted within the nebula, 
and re-scattered by neutral hydrogen. 
}
\end{abstract}

\begin{figure*}[h!]
\includegraphics[angle=0,width=120mm]{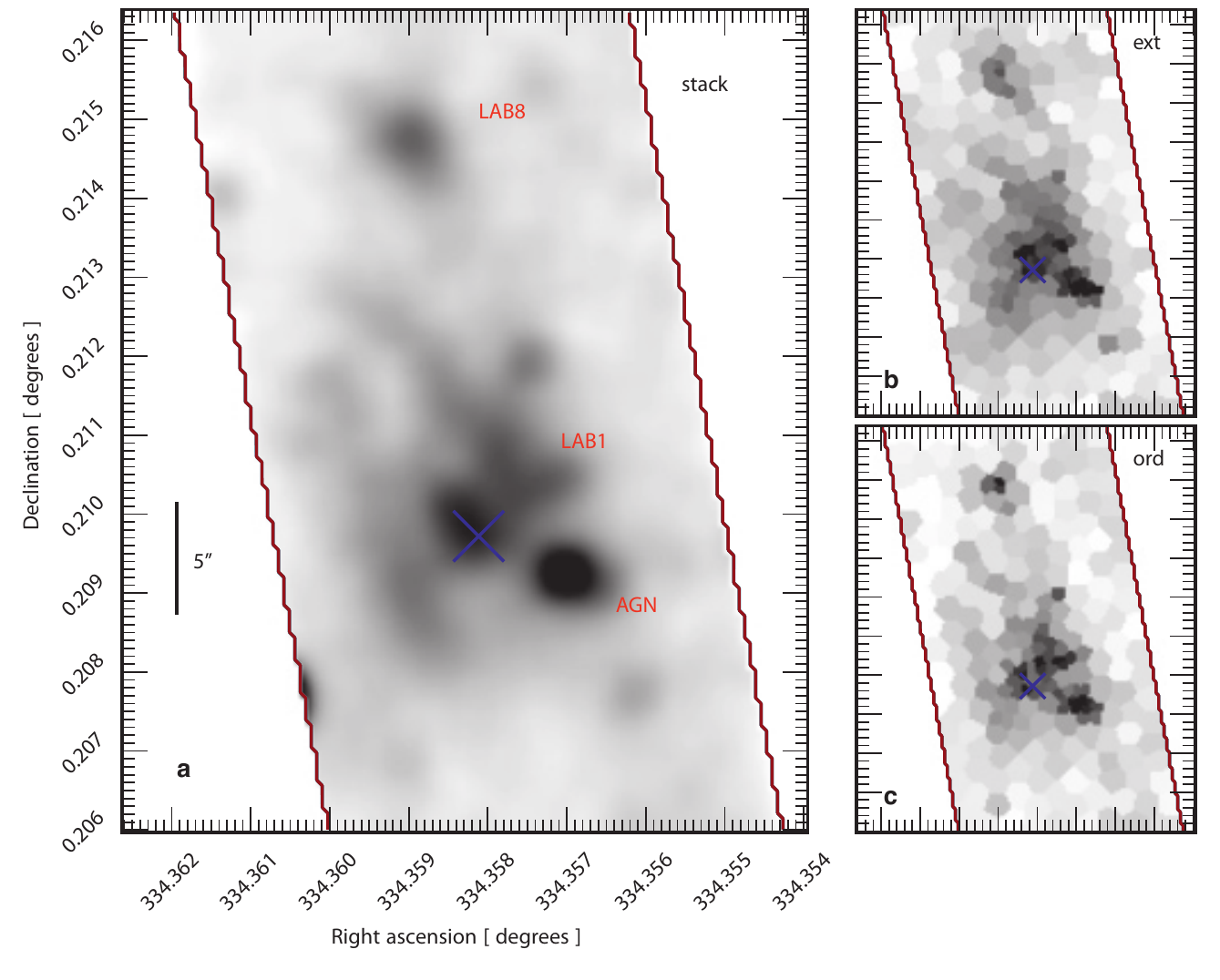}
\caption{~$\mathbf{|}$ \sffamily \textbf{LAB1 in \lya}. 
The large panel to the left shows the combined 
intensity frame that has been adaptively smoothed in order to show detail. 
The \lya\ emission is spatially extended over scales of more than 10 arcsec 
(80 kpc). The 5 arcsecond scale bar shown in the left panel corresponds to 40 
physical kilo-parsec. North is up and East is to the left. The primary target 
of this observation is LAB1 \cite{Steidel2000}, although the position angle
was chosen so as to also include LAB8 \cite{Matsuda2004}, which resides some
20 arcsec to the North. The bright point source to the South East (lower right)
is the AGN C11\cite{Steidel2000}. Major centres are labelled in the image. 
The two smaller frames to the right show the ordinary and 
extraordinary beams (lower and upper, respectively), which have been tessellated 
for quantitative analysis (Supplementary Information). All images are 
shown on the same intensity scaling, and even by 
eye differences in the intensity structure between the two beams become visible. 
The  cross marks the point that we determine to be the centre in the subsequent
analysis, and is determined as the point of peak \lya\ surface brightness that 
is not associated with the AGN. 
}
\label{fig:f1}
\end{figure*}

\begin{figure*}[t!]
\includegraphics[angle=0,width=120mm]{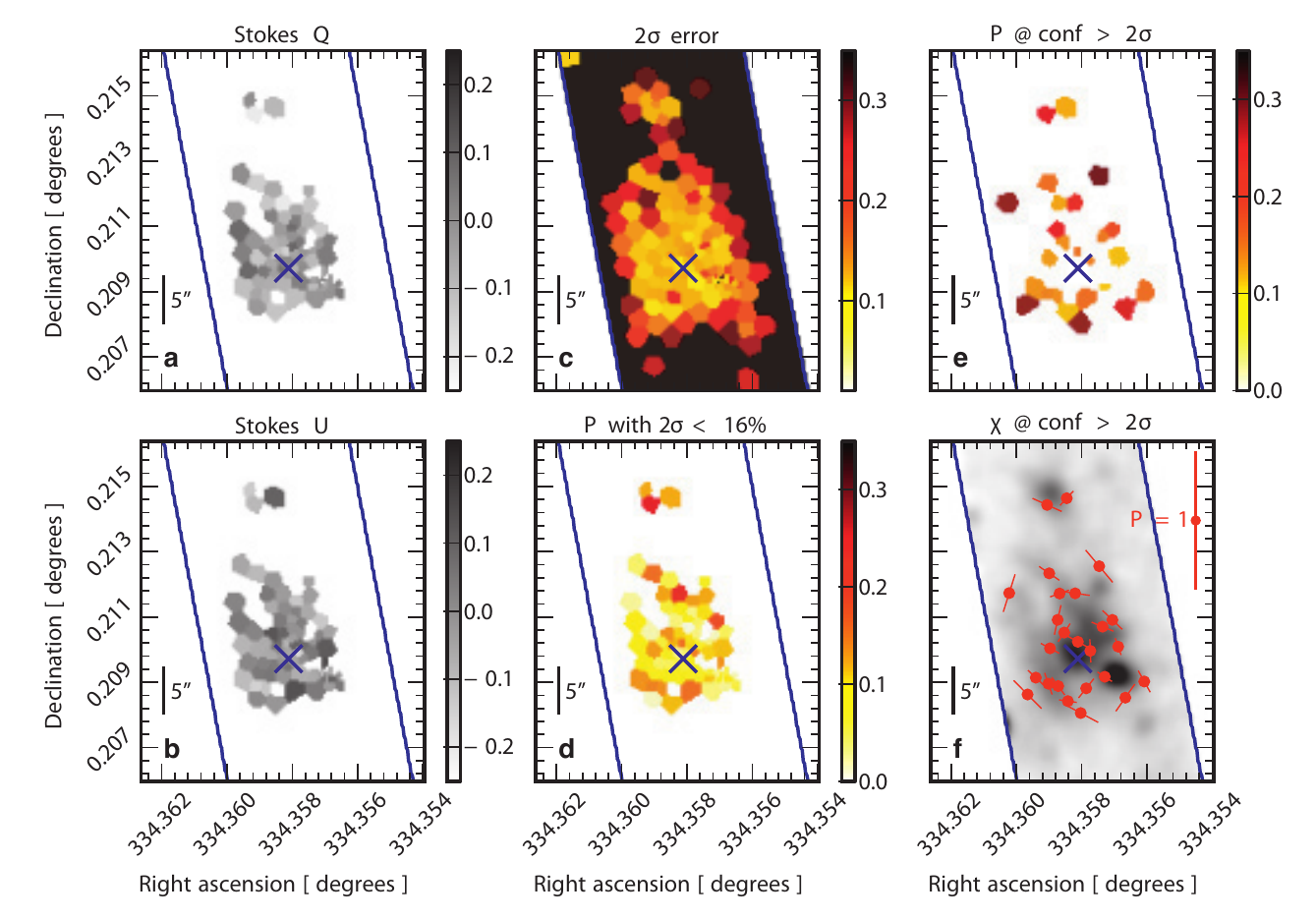}
\caption{~$\mathbf{|}$ \sffamily \textbf{The polarization of LAB1}. Scaling is shown to the right of each
panel. Panels (a) and (b) show the $Q$ and $U$ Stokes parameters. Panel (c) 
shows a map of the polarization fraction $P$ that we can measure at 95.4\% 
confidence; this error map demonstrates that in individual resolution
elements we can constrain $P$ in the central regions down to 10\% ($2\sigma$). At a radius 
of $\approx 6-7$~arcsec, this decreases to about the 20\% level, and beyond 
$\approx 10$~arcsec $P$ 
becomes noise-dominated. Panel (d) shows the value of $P$ in resolution elements
where it can be well constrained; we make a cut based on the error map, and show
bins where $P$ can be measured to better than 16\% ($2 \sigma$). $P$ of around 
5--10\% is found in the centre, but increases to 15--20\% at $\approx 5$ arcsec
from the centre towards the South, West, and North edges of LAB1. 
$P\approx 20$\%  is also seen around LAB8. Panel (e) shows $P$ in bins for which
we can be at least 95.4\% confident of its measurement. The structure around
LAB1 and LAB8 remains, extending to larger radii, and forming a broken ring at 
$P\sim 20$\%. 
Panel (f) shows a smoothed intensity image with the direction of the 
polarization vector, $\chi$, shown by red bars. As with panel (e) only bins
showing measurements at more than 95.4\% confidence level are shown. There is a
clear trend that, when bins show strong polarization, they orientate themselves
tangentially to the overall structure and local \lya\ surface brightness. These
features 
conform well with the predictions for the polarization of \lya\ photons that are
centrally produced and scatter at large radii, or are produced by cooling gas 
that flows onto the dark matter halo in quasi spherical symmetry
\cite{DijkstraLoeb2008}.  }
\label{fig:f2}
\end{figure*}

The \lya\ emission line of neutral hydrogen is a frequently used observational tracer of evolving galaxies
in the high redshift Universe. \lya\ imaging surveys 
typically find a large number of faint unresolved objects and a small
fraction of extremely luminous and spatially extended systems that are usually 
referred
to independently as Lyman alpha blobs (LABs). The compact sources usually appear to be more 
ordinary star forming galaxies whereas, since their discovery, much controversy 
has surrounded the true nature of LABs. 
Because one of the possible modes of powering LABs is the 
accretion of gas from the intergalactic 
medium\cite{Fardal2001,Yang2006,DijkstraLoeb2009,Goerdt2010,Faucher2010},
an eventual consensus about this powering mechanism holds particular relevance 
for how the most massive and rapidly evolving high redshift dark matter haloes 
acquire the gas required to fuel star formation in the galaxies they host. 
Predictions for polarized \lya\ radiation were made over a decade 
ago\cite{LeeAhn1998}, and have been further developed to the extent that 
the polarization signal may act as a diagnostic to study the origin 
of the \lya\ 
photons\cite{DijkstraLoeb2008}. By relying upon the scattering of \lya\ photons
at large distances from their production sites, these polarization measurements 
may also provide an independent
constraint on the distribution of neutral gas in the circumgalactic regions, 
which has proven a difficult measurement to make by traditional techniques. 
The one previous attempt to detect the polarization of a LAB returned a null 
result\cite{Prescott2011}, although very deep observations are required that 
need considerable time investments on 8--10 meter class telescopes. In the ESO
Very Large Telescope observations presented here, we observe one of the largest
and most luminous LABs on the sky in imaging polarimetry mode 
(Figs~\ref{fig:f1} and \ref{fig:f2}).

These new, deep polarimetric observations show a substantial fraction of the 
\lya\ emission from LAB1 to be polarized. The data enable the 
fraction of light that is linearly polarized ($P$) to be measured to better than
5\% (r.m.s.) in the regions of peak surface brightness, however in these central regions
we find essentially no polarization signal. In order to detect the polarized 
emission we need to examine the fainter signal emanating a few arcsec from the centre --
when \lya\ is emitted with a high polarization fraction ($P \approx 20$\%), it is not 
spatially coincident with the brighter regions of the nebula.
Indeed the highest and most statistically certain
measures of $P$ are found some 4--8 arcsec from the centre, forming an almost complete
ring of polarized emission. 
Just a single resolution element shows a signal-to-noise ratio on $P$ that exceeds 3,
but our confidence in these results is substantially enhanced by the strong degree 
of spatial correlation between the resolution elements that are observed to be polarized 
at high significance. 
Away from the regions of bright \lya\ emission our data show no measurable polarization 
signal, but a second nearby extended \lya\ source 
(LAB8\cite{Matsuda2004}) that lies to the north also shows strongly polarized \lya.

\lya\ is a resonant line and photons scatter in neutral hydrogen (H~{\sc i}),
undergoing a complicated radiation transfer. At each scattering event,
of which there may be millions, a photon acquires a new direction and
frequency, dependent upon the kinematics of the scattering medium and quantum mechanical
probabilities\cite{Stenflo1980}. The outgoing frequency determines the optical depth
that the photon then sees\cite{Adams1972,Neufeld1990}:
small shifts from line-centre (core scatterings) result in effectively
no transfer, while long frequency excursions (wing scatterings) result
in significant flights and likely escape from the system. 
The more frequent core scatterings may emit polarized radiation but it is typically with a 
low fraction ($P\lesssim 7$\%); the comparatively
rare wing scatterings, on the other hand, may introduce a high degree of 
linear 
polarization to the outgoing radiation ($P$ up to 40\%)\cite{Stenflo1980,DijkstraLoeb2008}. 
The observed polarization
signal therefore provides an unambiguous signature of the scattering
nature of the diffuse \lya\ halos, and distribution of circumgalactic gas.  
On the atomic level, the detected polarization signal provides
evidence that the \lya\ photons are not emerging after
a large number of resonant core scattering events, but are exiting after
long shifts in frequency.

The detection of polarization demonstrates that the \lya\ photons cannot have 
been produced in situ at large radial distances. In order to impart a significant level of
polarization on the outgoing radiation, a geometry is required in which \lya\
photons preferentially escape after scattering off hydrogen atoms at angles of 
90 degrees to their vector prior to the scattering event\cite{Stenflo1980,DijkstraLoeb2008}. 
This geometry is not expected in the case of wind/shock excitation since
thermalized cooling gas will generate \lya\ photons with no preferential 
orientation or impact angle with respect to the scattering medium. Similarly, the 
filamentary streams of cold infalling gas predicted by simulation\cite{Keres2005,Dekel2008}
appear to have too small volume filling factors, and be too narrow (few kpc) for any 
imparted polarization signal to be observable\cite{DijkstraLoeb2009} -- 
a spatial resolution would be required that is currently unattainable by 
optical polarimeters. While 
neither of these rejected scenarios have yet been thoroughly tested by 
radiative transport studies, the production of polarized \lya\ is well understood, 
and it is difficult to see a mechanism by which observably polarized \lya\ 
may be produced in these situations.
We therefore conclude that the \lya\ photons must have been produced centrally in the galaxies
themselves and scattered 
at large radii by neutral gas. Even if the halo is accreting intergalactic gas,
this process is not responsible for a large fraction of the \lya\ emission. 

The geometrical origin of the polarization signal in the photon scattering model, 
which is analogous to Rayleigh scattering, implies also that the angle of the 
polarization vector $\chi$ should align tangentially to the overall geometry of the system
at large radii\cite{RybickiLoeb1999,DijkstraLoeb2008}.
We also extract this $\chi$ vector from the data (Fig~\ref{fig:f2}, panel f). If we 
define the geometric centre to be the region of peak \lya\ surface brightness (excluding the 
active nucleus; Fig 1), these vectors display a strong tendency to orientate 
along directions tangential to the overall geometry of the system. Again this is 
in good agreement with 
theoretical expectations, should the \lya\ photons be produced centrally by ordinary star-formation
and/or AGN-fueled processes and 
scatter at large radii\cite{DijkstraLoeb2008}. However unlike the idealized models
used for radiative transport simulations, which assume
perfect spherical symmetry and no velocity shear or clumping of the gas, the true
\lya\ surface brightness is far from smooth and symmetric. This may reflect 
either multiple
\lya\ sources within the halo, large scale inhomogeneities in the scattering medium,
or both. Polarization of \lya\ emission in a clumpy medium has not yet been 
tested by simulation although clumping could potentially either suppress or enhance the 
polarization\cite{DijkstraLoeb2008}, and likely depends on the relative size and
volume filling factor of the clumps. 
Regardless, we would still expect the polarization signature to remain
provided that the individual regions are well resolved, in which case 
$\chi$ would then be expected to align tangentially to contours in the local surface 
brightness. This effect is also reflected in the data, and we find polarization vectors 
preferentially line up at angles approximately perpendicular to the local surface brightness 
gradient at their position. 
The Supplementary Information includes a 
quantitative analysis of the polarization angles compared with both local surface 
brightness gradients and the overall system geometry.

\begin{figure*}
\includegraphics[angle=0,width=120mm]{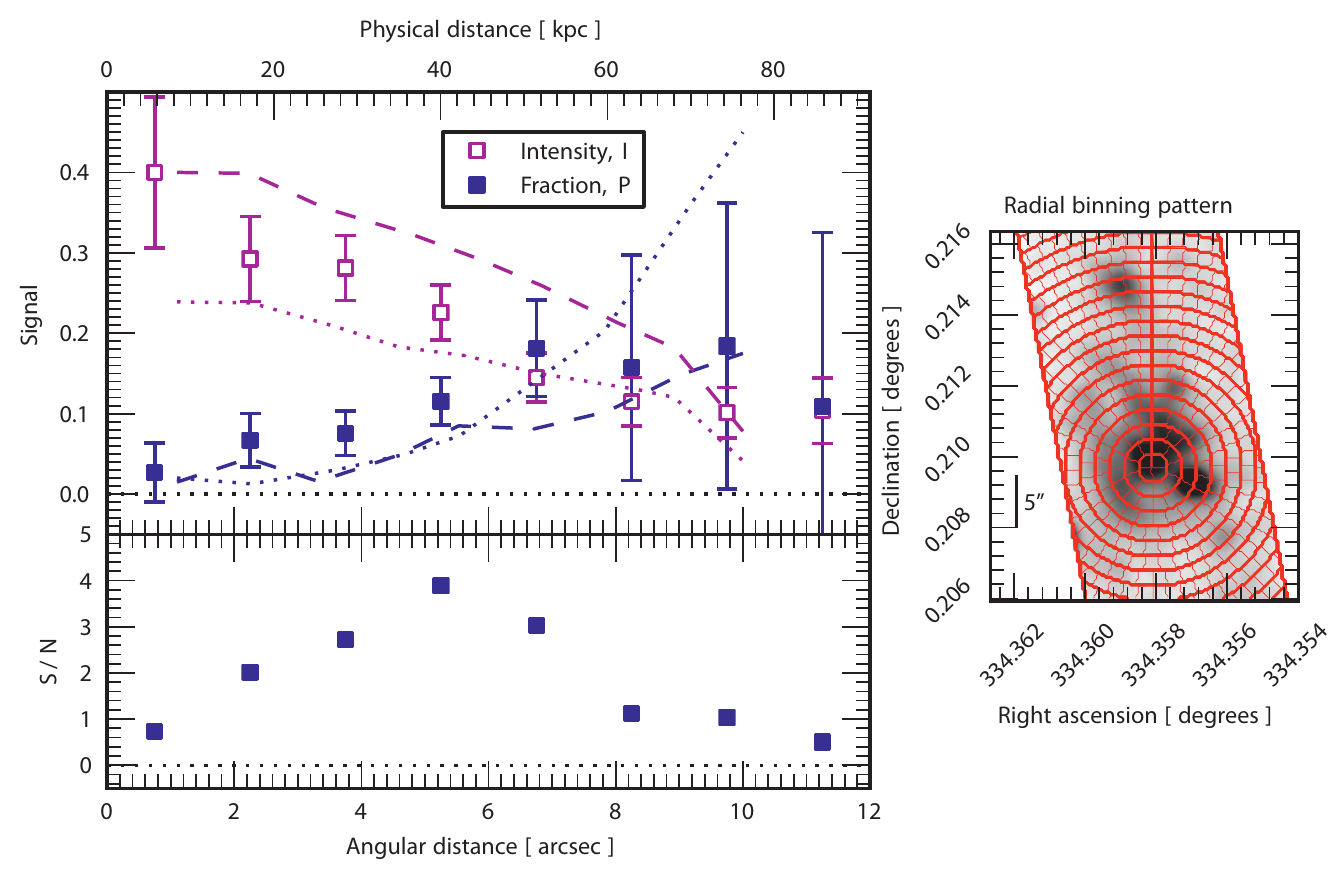}
\caption{~$\mathbf{|}$ \sffamily \textbf{The radial polarization profile}. Blue points
show the fractional polarization, $P$, which is measured in concentric 
1.5~arcsec annuli. We implement a new binning method (Supplementary Information)
that enables annular measurements of $P$ using independent bins. 
The binning structure is shown
to the right, the centre the same as illustrated in 
Fig~\ref{fig:f1}. $P$ has been corrected for 
spuriously high polarization fractions that may be measured at low 
signal-to-noise\cite{WardleKronberg1974}. In magenta the total intensity profile
is shown. Errorbars on both measurements are $1\sigma$, and 
the lower profile shows the signal-to-noise ratio of $P$. In the central 
regions where intensity is highest, $P$ is 
measured at $\sim 4$\%, although at $1\sigma$ is consistent with no 
polarization. $P$ and its signal-to-noise both increase with radius; by 5 arcsec
$P=10$\% and is inconsistent with zero at the $4 \sigma$ level. $P$ peaks at 
18\% at 7~arcsec, beyond which radius the noise becomes dominant. 
Overlaid are the results of numerical models for scattered \lya\ 
radiation\cite{DijkstraLoeb2008} in an expanding \hi\ shell with two column 
densities: $10^{19}$~cm$^{-2}$ (lower) and $10^{20}$~cm$^{-2}$ (upper).
The curves have been smoothed over lengths of 1.5 arcsec to match our 
annuli.  
The actual radius
of the shell is unknown, so the theoretical points have been scaled to shell radius 
of 10 arcsec,
in order to approximately match both the surface brightness and polarization 
profiles. This shows that the predictions are qualitatively matched by observation.
Models for scattering in a static IGM and spherically symmetric
accretion are not show since they are not leading explanations for the powering
of LABs.  }
\label{fig:f3}
\end{figure*}

There is one final observable prediction that has been suggested by 
radiative transport simulations. This is the radial profile of the polarization fraction,
$P$, which according to the scattering shell model should present with $P\approx 0$ in the
geometric centre and increase with increasing radius.
We have computed this radial profile of $P$ using the same central coordinates as defined previously
(Fig~\ref{fig:f3}) and find strong evidence for $P$ increasing with radius, again in 
agreement with predictions. 
The central point on the profile and most significantly polarized point in the ring 
are discrepant at over $2\sigma$. 
We have fitted the innermost 7 arcsec (5 data points; where data are deemed reliable) 
with the simplest 
function for $P$ varying linearly with radius. Computing the classical $p-$value, 
$1-\Gamma(\nu/2, \chi^2/2)$ where $\nu$ is the number of degrees of freedom, shows above 96\% confidence in this hypothesis.  
The same test performed on a flat profile ($P=8.3$\%), suggests that 
a structure in which $P$ is independent of 
radius is consistent with the data at 16\%. Regardless, the slim possibility
of $P$ that is flat with radius 
does not present a difficulty for the scattering scenario, which is supported not only 
by the detection of polarized \lya\ itself, but also by the angular distribution of 
polarization vectors; this radial profile is simply the least robust of our constraints.
The actual gradient of $P$ with radius
depends upon both the column density and radius of the shell, neither of which 
we know -- assuming that the idealized geometry holds and that the shell has a radius of 10 arcsec, 
the models with column densities of $10^{19}-10^{20}$~cm$^{-2}$
would be in agreement with the data (Fig 3). 
In this step of the analysis we also compute the total luminosity-weighted polarization 
fraction within each radius: within 7 arcsec we obtain an 
integrated value of $P=11.9\pm 2$\%. We take this value as the overall 
polarization fraction of the nebula, which we quote at $6\sigma$ significance.

\begin{addendum}
\item Based on observations made with ESO Telescopes at the Paranal Observatory under
programme ID 084.A-0954.
M.H. was supported in part by the Swiss National Science Foundation, and also
received support from Agence Nationale de la recherche bearing the reference
ANR-09-BLAN-0234-01.
Patrick Ogle is thanked for his valuable suggestions about polarimetric
observations. We are indebted to Nino Panagia and Terry Jones for comments on 
the manuscript, and to Daniel Schaerer, Nick Scoville, Chris Lidman, Arjun Dey,
Moire Prescott, and Paul Lynam for discussions. 
\item [Author Contributions] All authors contributed to the proposal preparation. M.H. and C.S.
observed and reduced the data. M.H. analyzed the results.  All authors contributed to the manuscript
preparation.
\item [Competing Interests] The authors declare that they have no
competing financial interests.
\item[Correspondence] Correspondence and requests for materials
should be addressed to M.H.\\
(email: matthew.hayes@ast.obs-mip.fr).
\end{addendum}

\clearpage

\end{document}


\title{Supplementary Information}

\maketitle

\section{Strategy and Observation}

For the first deep polarimetric study of a \lya\ blob there were just several
tens of objects on the sky from which we could have selected our target. 
According to predictions of the structure of the 
polarization$^{17}$
the selected target needs to be azimuthally 
resolved in order to make a detection: regions with a given angle of 
polarization, $\chi$, must not all be superimposed under one (or a few) seeing 
discs. The result of this would be a collapsed vector sum of zero. Furthermore 
an appropriate target needs to have sufficient 
surface brightness at radius of a few arcsec (preferably at least 5) to examine the 
radial profile of polarization, and also simply to minimize the 
observation time on an exploratory and relatively costly first-time observation. 
Finally, analysis benefits greatly from selecting a well-studied object with as 
much ancillary data as possible to further constrain the physics of the blob
and its counterpart(s). The object that was deemed to best satisfy these 
criteria was LAB1 ($\alpha=22^h17^m25.^s7$; 
$\delta=+00^d12^m49^s.6$, J2000$^2$, located in the
SSA22 protocluster at $z=3.09$.

\begin{figure*}
\includegraphics[angle=0,width=80mm]{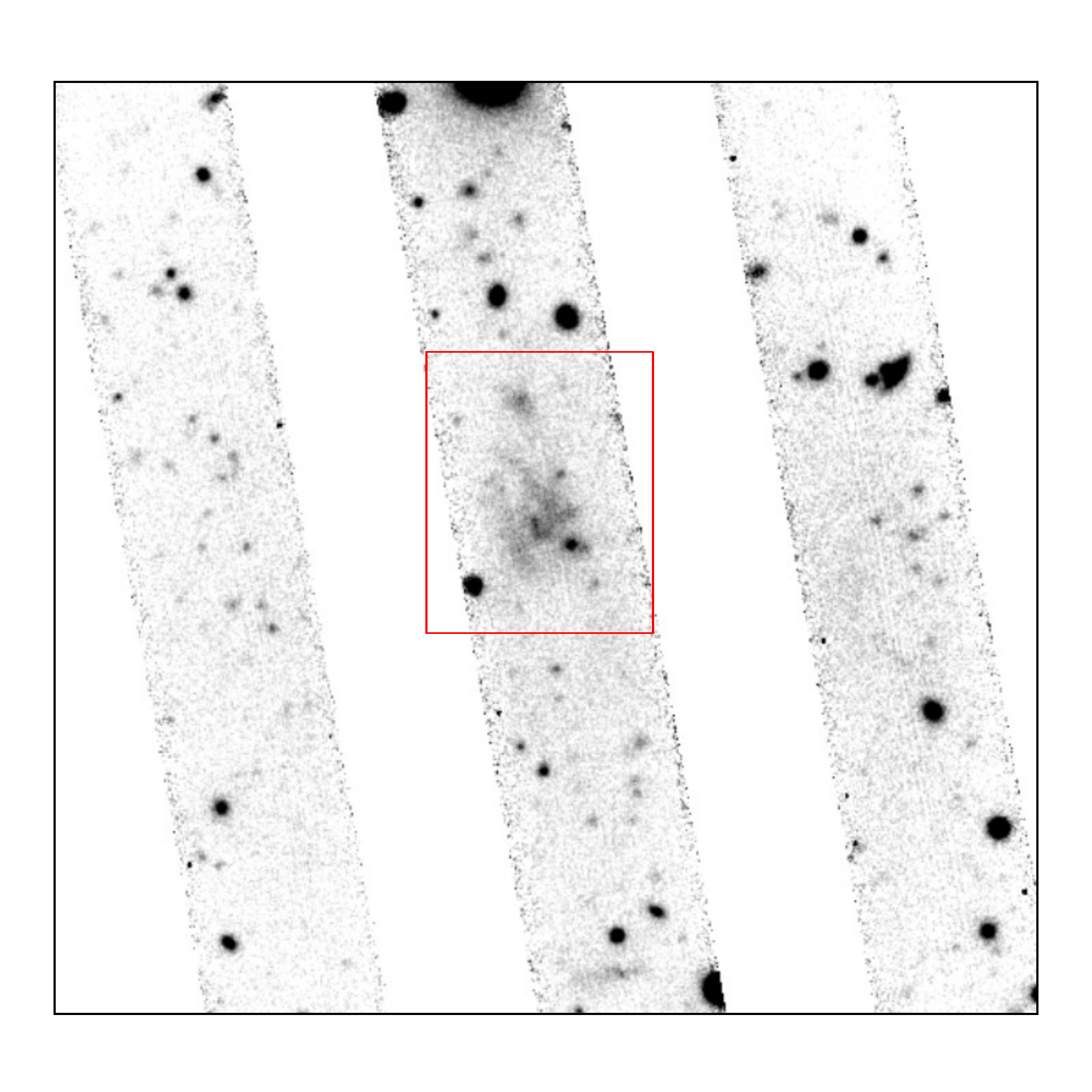}
\caption{~$\mathbf{|}$ \sffamily \textbf{LAB1 in the SSA22 protocluster as seen by FORS2}.
The field is split
into parallel stripes that are re-projected after splitting into orthogonally 
polarized ordinary and extraordinary beams. Shown here is the ordinary beam, so 
these stripes show only 50 per cent of the frame. The red box shows the region 
of the field cut out for Figures~1 and 2 of the main article.}
\label{fig:rawfield}
\end{figure*}

LAB1 was observed during the first half of four consecutive nights beginning 
20-23 October 2009, using the FOcal Reducer and low dispersion 
Spectrograph\cite{Appenzeller1998}  (FORS2) instrument mounted in
the Antu (UT1) node of the Very Large Telescope at the European Southern 
Observatory. 
The first stage in the dedicated polarization optics of FORS2 is the 
introduction of nine evenly spaced 22~arcsecond wide MOS strip masks, which 
split the incoming field into evenly separated stripes, removing half of the 
field. The light is then passed through a super-achromatic half-wave plate 
(HWP) retarder mosaic ({\tt RETA2+5}), which rotates the angle of polarized light. 
Four position angles of the HWP plate are required for unambiguous measurements
of the $Q$ and $U$ Stokes parameters, which simultaneously suppresses the 
impact of possible improper flat-fielding and sky subtraction (see below). 
We adopt the standard angles for the recovery of $Q$ and $U$: 0\degree, 
22.5\degree, 45\degree, and 67.5\degree.
The rotated beam is subsequently passed through a Wollaston prism 
({\tt WOLL\_34+13}), which splits randomly polarized or unpolarized light into
two orthogonal, linearly polarized outgoing beams; arbitrarily denoted as the 
`ordinary' ({\tt ord}) and `extraordinary' ({\tt ext}) beams. These beams are 
subsequently re-projected onto the CCD, duplicating the 22~arcsecond beams from
the MOS mask as simultaneously observed {\tt ord} and {\tt ext} images 
(illustrated in Figure~\ref{fig:rawfield}).
Finally, in order to isolate the $z=3.1$ redshifted \lya\ line, we used a 
narrowband [O {\sc iii}] filter designated {\tt OIII+50}, with a central 
wavelength of 5,001~\AA\ and full width half maximum of 57~\AA.

Throughout the course of each night the sequence of four HWP angles was observed
repeatedly, using integration times of 1,200~seconds at each position.
This 
sequence was repeated three times on the first three nights, and twice on the 
last night (Oct 23).  Thus at each retarder position angle we obtain a total
integration time of 13,200~s. 

FORS2 is known to introduce a spurious instrumental polarization 
signature although this signature is small, contributing a fraction of 
polarization, $P$, of $\lesssim 3\times 10^{-4}$ ($V-$band) in the centre of 
the field of view, and increases smoothly with radius. 
We do adopt a small dithering procedure and also step the target between
stripes on the MOS mask, which requires steps of 44 arcsec in the $y-$direction 
of the chip. However the object is always positioned 
within 2 arcmin of the centre, at which 
radius the instrumental signal will contribute $P$ of 0.2 percent or less. 
Unlike typical polarization studies of either stars of 
point-like sources, we are working with faint, diffuse radiation
and within the very best limits of the data we can achieve a statistical error 
on $P$ of no better than about 5 percent. Thus any systematic at the 0.2\% 
level will have a negligible impact upon the measurement on target. 

The bulk of the observing time was classified as clear although over half 
was photometric. However, the starting point in the polarization calculations 
is the ratio of ({\tt ord}--{\tt ext})/({\tt ord}+{\tt ext}) at each position of
the HWP (Equation~\ref{eq:normdiff}). 
Since the {\tt ord} and {\tt ext} beams are obtained simultaneously, deviations 
from photometricity have no impact upon the 
determination of the Stokes parameters. Astronomical seeing varied only 
moderately over the 20 hours of science time, between 0.85
and 1.15~arcseconds.

Observations were carried out close to new moon in order to minimize the sky 
background at bluer wavelengths. On nights beginning 20, 21, and 22 the moon 
was anyway below the horizon during the course of the observations, but rose 
on the night beginning 23 Oct, approximately doubling the sky background 
intensity compared to the first night.

Using the same setup as described above, observations of one highly polarized 
spectro-polarimetric standard star and one unpolarized standard 
were also obtained.
The polarized star was BD-12\,5133 ($\alpha=$18:40:02; $\delta=$--12:24:07; 
$P=4.32\pm 0.05$\%; $\theta=148.4 \pm 0.3$\degree) and the unpolarized star
WD\,2149+021 ($\alpha=$21:52:23; $\delta=$+02:23:07). More information can be 
found on the ESO webpages for the FORS polarimeter\footnote{
\href{http://www.eso.org/sci/facilities/paranal/instruments/fors/inst/pola.html}{http://www.eso.org/sci/facilities/paranal/instruments/fors/inst/pola.html}}. 
Integration times were set to 2~seconds for both stars and the sequence of four
angles was repeated twice at each visit. Both standard stars were visited once 
per night on nights beginning 20, 21, and 22, while no standards were obtained on
night beginning Oct 23.

\section{Data reduction}

The initial steps of data reduction were carried out in the standard manner for
imaging observations (i.e. without polarizing optics) using 
{\tt NOAO/IRAF}\footnote{IRAF is distributed by the National Optical Astronomy 
Observatories, which are operated by the Association of Universities for 
Research in Astronomy, Inc., under cooperative agreement with the National
Science Foundation.}. 
Individual frames were bias subtracted and corrected for pixel-to-pixel 
variations using well exposed twilight flats. Even though the MOS striped images are 
re-projected after beam separation into simultaneous {\tt ord} and {\tt ext}  
beams, traditional flat fielding methods were applied. Ideal flat fielding would 
incorporate the
systematic effects of all the polarization optics, including the {\tt ord} and 
{\tt ext} beam splitting, but this is not practical for most setups of imaging 
polarimetry\cite{Patat2006}: local pixel-to-pixel 
variations are anyway accounted for by traditional flat-fielding and the effect
of imprecise relative flat-fielding between the two beams is accounted for by
the redundancy against the four HWP angles: with four angles, swapping beams 
{\tt ord} and {\tt ext} by $\pm \pi/4$ gives a factor of 2 redundancy in Stokes 
parameter sampling
[ ({\tt ord}, 0\degree) = ({\tt ext}, 45\degree), 
	({\tt ext}, 22.5\degree) = ({\tt ord}, 67.5\degree), etc. ].

Individual frames were all corrected for atmospheric extinction using an 
extinction coefficient of 0.1412 and the airmass at the midpoint of each 
observation. The coefficient was derived by linearly interpolating 
the extinction coefficients of the FORS2 {\tt b\_HIGH} and {\tt v\_HIGH} 
filters (October 2009 -- March 2010, from the FORS2 quality control webpages)
to 5,001\AA.

Cosmic rays were coarsely removed by using the {\tt crutil.crmedian} task in 
{\tt IRAF} for two reasons: firstly to prevent them from contaminating flux
measurements of the local sky background, and secondly to prevent them from
interfering with centroid measurements of the stars used for image registration. 
I.e. the cosmic-ray rejected frames were not used in the final image stacks 
used for science, for which the original (cosmic-ray infected) images were 
stacked after the sky surface and image alignment solutions had been found
(coming paragraphs). 

In the next stages, all of the individual frames  were split 
into their individual {\tt ord} and {\tt ext}  beams. All images 
($11\times 8=88$ in total) were registered to a common astrometric grid using 
the {\tt geomap} and {\tt geotran} tasks of {\tt IRAF}. A {\tt general} transform 
geometry was adopted throughout, which provided an r.m.s. of less than 1 pixel in
all cases, and also to removed a slight geometrical distortion that is  
introduced between the {\tt ord} and {\tt ext} beams by the polarizing optics 
(consistently $\sim 1.2$ pixels in y- r.m.s. when computed with a simple shift and 
 rotation). 

A strong 
centrally concentrated sky-like structure is seen in each of the reduced frames,
which is a well-known FORS instrumental feature
\footnote{\href{http://www.eso.org/~fpatat/fors/polarimetry/polarization.html}{{\tt http://www.eso.org/\~{}fpatat/fors/polarimetry/polarization.html}}}. 
This was removed in the same step as common sky subtraction by fitting a 
two-dimensional  polynomial surface using the {\tt fit/flat\_sky} 
task {\tt ESO/MIDAS}. 
All astrophysical sources were avoided in the determination of the sky 
background, which was estimated from images in which cosmic rays had been
removed. Boxes in which the sky was estimated used over 30\% of the total 
pixels in the science frames, and the same sky boxes were used for all frames.
Each background surface frame was saved and checked for general consistency 
with its neighbours by visual inspection. Signal-to-noise in a sky measurement
always exceeded 100, even in the smallest boxes. 
Once the coordinate transforms and sky backgrounds had been computed, the 
individually reduced frames ({\it before} cosmic ray rejection) were 
background subtracted and shifted, returning the cosmic rays to the aligned, 
sky-subtracted frames. Images were then co-added using a mean combination with 
simple {\tt minmax} rejection of the highest and lowest two values at each pixel using
{\tt imcombine}. The $1\sigma$ standard deviation map was stored. In total we 
combine eight groups of 11 images to produce individual science frames for 
polarimetry ({\tt ord} and {\tt ext} at the four angles), 
four groups of 22 images to produce the individual intensity images at each
angle ($I=${\tt ord+ext}),
and one group of all the 88 science frames to give good $S/N$ in a frame used 
for aesthetic purposes. 
Seeing in the final frames was measured to be 0.96~arcseconds.

Standard stars were reduced in an identical manner using the same scripts.

\section{Binning of the science frames}

\subsection{Voronoi Tessellation}

Examining the stacked science frames and their associated r.m.s. maps, we see
that in the very brightest regions of LAB1, the signal-to-noise ratio 
$(S/N)$ in individual pixels is around 2. Thus for the vast majority of the 
area of interest, surface brightness measurement will have $S/N$ per pixel 
substantially below 1. The signal-to-noise on $P$, $(S/N)_P$, is 
given by $(S/N)_P = \sqrt{N_\mathrm{HWP}/2} \cdot P \cdot (S/N)_\theta$,
where $N_\mathrm{HWP}$ is the number of HWP angles, and $(S/N)_\theta$ is 
simply $S/N$ in the individual intensity frames\cite{Patat2006}. These 
observations were designed to give us $(S/N)_P=3$ for $P=20$\% at a radial 
distance of $\sim 5$~arcseconds from the centre:
in order to obtain this we need 
$S/N_\theta = 3 / 0.2 /\sqrt{4/2} \approx 10$ per spatial 
resolution element in the individual frames.

It is therefore clear that we need to perform some binning in our science 
frames before we compute the Stokes parameters. However, the \lya\ 
polarization signal results from the geometrical configuration of the 
H~{\sc i} gas, and the maps must be azimuthally resolved in order to not 
smear together the information, giving Stokes parameters averaged to zero. 
In short: some binning is necessary to estimate $P$, while over-binning will make 
it unmeasurable. Here we are 
fortunate that in the central regions $S/N$ is anyway substantially higher. 
In order to bin our data we adopt the method of Voronoi tessellation which 
dynamically accretes pixels together according to the local surface brightness 
until a threshold $S/N$ has been obtained in a given bin. To this end 
we made use of the Weighted Voronoi Tessellation (WVT)
algorithm\cite{DiehlStatler2006}, which is a generalization of the 
Voronoi binning algorithm\cite{CappellariCopin2003}. 

We first operate on the combined image stack of both {\tt ord} and {\tt ext} 
beams taken at 0\degree. By working on this combined frame we do not bias our 
binning structure towards certain vectors of polarization. We begin the 
procedure at the 
brightest pixel in the central region (marked in Figure 2 of the main article), 
target a threshold $(S/N)_\theta= 10$,
and enforce a maximum bin size of $8\times 8$ pixels or $4$~square arcseconds.
After generation of the binning pattern we then apply the same structural 
binning to all of the eight individual science frames, and save the binned 
signal and noise maps. The pattern is best illustrated in Figure~1 of the main article.

\subsection{Radial Binning}

In order to produce a radial map of $P$
we are less interested in preserving local details to examine
the polarized morphology, and are concerned only with getting an average at 
each radius. Here we cannot simply operate on a polarization map derived from 
the unbinned science 
frames because of signal-to-noise issues and the error distribution of $P$:
any pixel for which
the intensity is significantly affected by local noise will cause $P$ to skew 
to very high values (often above 100\%). 
We cannot perform a radial analysis on the Voronoi binned images because individual bins may 
contribute to more than one annulus, which will correlate the errors. We need a
new binning method in which each pixel contributes exactly once to a bin, and 
each bin contributes exactly once to an annulus.

We adopt a new
binning method: moving outwards from the centre in uniform concentric annuli, we
sub-bin each ring in even angular steps. Even sub-binning is required for 
straightforward averaging of bins and error combination within an
annulus. We employ 1.5 arcsec (6 pixel) annuli and divide each annulus into bins
of 36 pixels, which corresponds to boxes of about $1.5 \times 1.5$ arcsec. In 
practice these bins actually contain between 35 and 38 pixels due to angular 
effects at the edges, but we ignore this minor source of imbalance.

\section{Polarization Calculations and Monte Carlo}

For linear polarization, the decomposition of polarized light falls only into 
the $Q$ and $U$ normalized Stokes parameters. The $V$ parameter represents 
circular polarization and is anyway not expected for \lya\ radiation, and is 
subsequently ignored.
The fourth parameter $I$ is the total intensity: the flux measured by narrowband
imaging, which is the sum of the {\tt ord}+{\tt ext} beams. 
For each HWP position, $\theta$, the normalized flux difference, 
$F_\theta$, is defined as: 
\begin{equation}
F_\theta = \frac{f^\mathrm{ord}_\theta - f^\mathrm{ext}_\theta}{f^\mathrm{ord}_\theta + f^\mathrm{ext}_\theta}
\label{eq:normdiff}
\end{equation}
In the case where four HWP angles have been obtained, $Q$, $U$, and $I$
relate to the observables by: 
\begin{eqnarray}
\bar Q \,\,\, & = & \,\,\, \frac{Q}{I} \,\,\, = \,\,\, \frac{1}{2} \cdot F_{0.0} \,\,  - \frac{1}{2} \cdot F_{45.0} \nonumber  \\
\bar U \,\,\, & = & \,\,\, \frac{U}{I} \,\,\, = \,\,\, \frac{1}{2} \cdot F_{22.5} - \frac{1}{2} \cdot F_{67.5} 
\label{eq:stokes}
\end{eqnarray}
From these, the polarization fraction $(P)$ and polarization angle $(\chi)$ 
follow as 
\begin{eqnarray}
P    \,\,\, & = & \sqrt{\bar Q^2 + \bar U^2} \nonumber  \\
\chi \,\,\, & = & \frac{1}{2} \cdot \arctan \frac{U}{Q}
\label{eq:polar}
\end{eqnarray}

We used Equations~\ref{eq:normdiff}, ~\ref{eq:stokes}, and \ref{eq:polar} to 
compute the $P$ and $\chi$. Formal errors are computed by Monte Carlo methods
which have been shown to be consistent with analytical error propagation
at high signal-to-noise\cite{Patat2006}, and are certainly more 
appropriate to handle the low
signal-to-noise data (by polarimetry standards) presented here. When running a
Monte Carlo simulation we perform 10,000 realizations and assume that the input 
errors follow Gaussian deviates, which is a realistic assumption 
for photometry based on modern optical CCDs.

While the input deviates are Gaussian, the error distribution on $P$ that 
follows from Equations~\ref{eq:normdiff}, 
\ref{eq:stokes}, and \ref{eq:polar} will be far from normal.
Equation~\ref{eq:normdiff} shows that the distribution of normalized 
differences, $F_\theta$, will already take a positive skewness since they
are the ratios of fluxes: if the denominator is a symmetric Gaussian 
distribution it will skew the total distribution of $F_\theta$ positively.  Furthermore, 
Equation~\ref{eq:polar} shows $P$ to be the quadrature sum of $Q$ and
$U$, and hence the distribution of $P$ will be further positively skewed
(see Figure~\ref{fig:disthist} for an illustration). Thus the only route we have 
to the confidence limits on our measurements comes through analysis of the 
probability density function (PDF) of all the Monte Carlo realizations. 
We compute the integrated PDF and determine the lower and upper 
confidence limits on $P$ that correspond to 1, 2 and $3\sigma$. 
Using the measured value of $P$ (I.e. in the single,
non-perturbed raw data without Monte Carlo), 
we conservatively define the formal $N\sigma$ error as the larger of the two 
absolute deviations from the measurement. 
When working purely in the noise with no genuine signal, no
orthogonality can be expected between the $Q$ and $U$ vectors, and the quantity 
$P$ may mathematically exceed 100\%, although the significance of this will be zero.
Illustrations of the error calculation are shown in Figure~\ref{fig:disthist}.

\subsection{Data Calibration and Polarization Measurements}

\begin{figure*}
\includegraphics[angle=0,width=54mm]{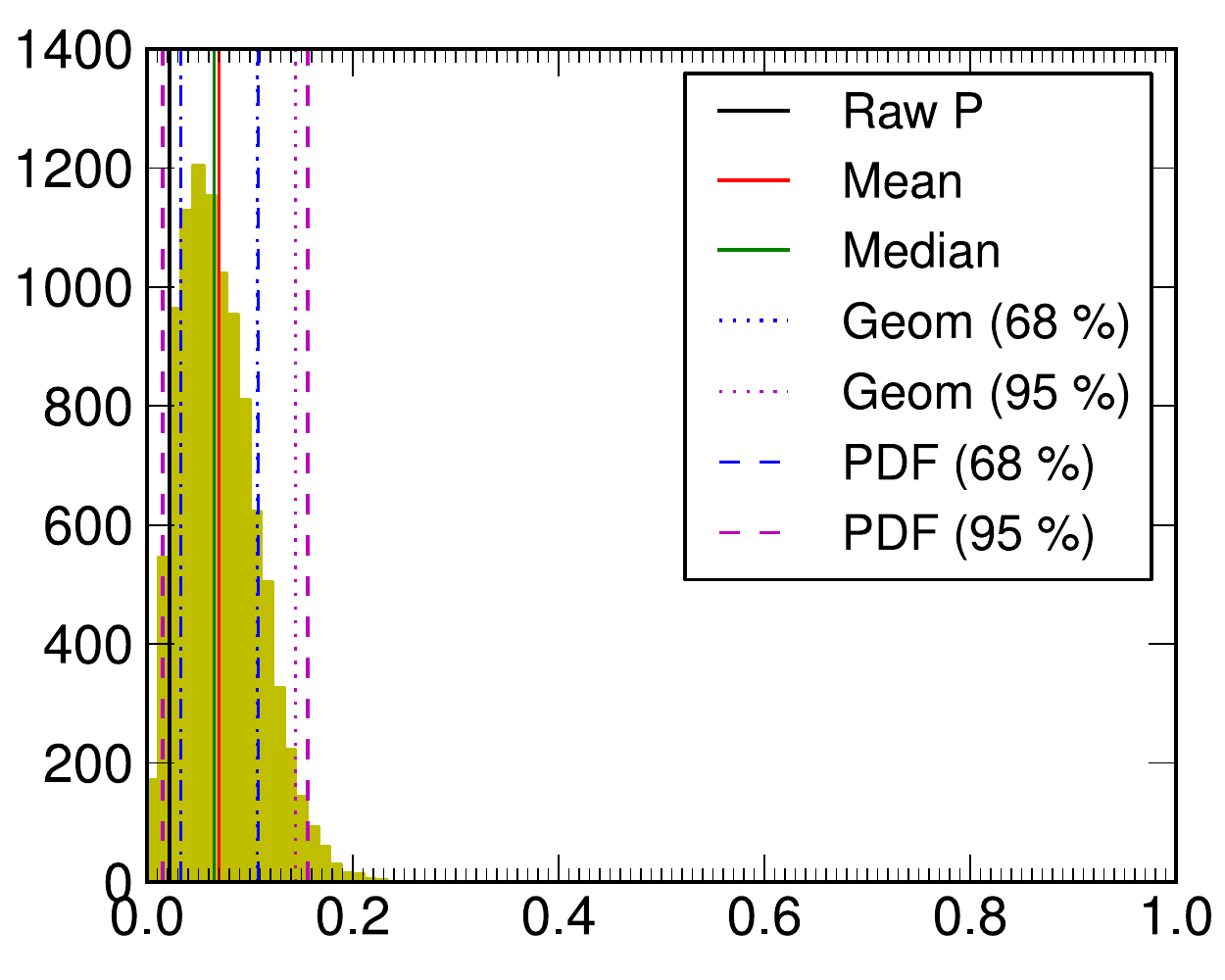}
\includegraphics[angle=0,width=54mm]{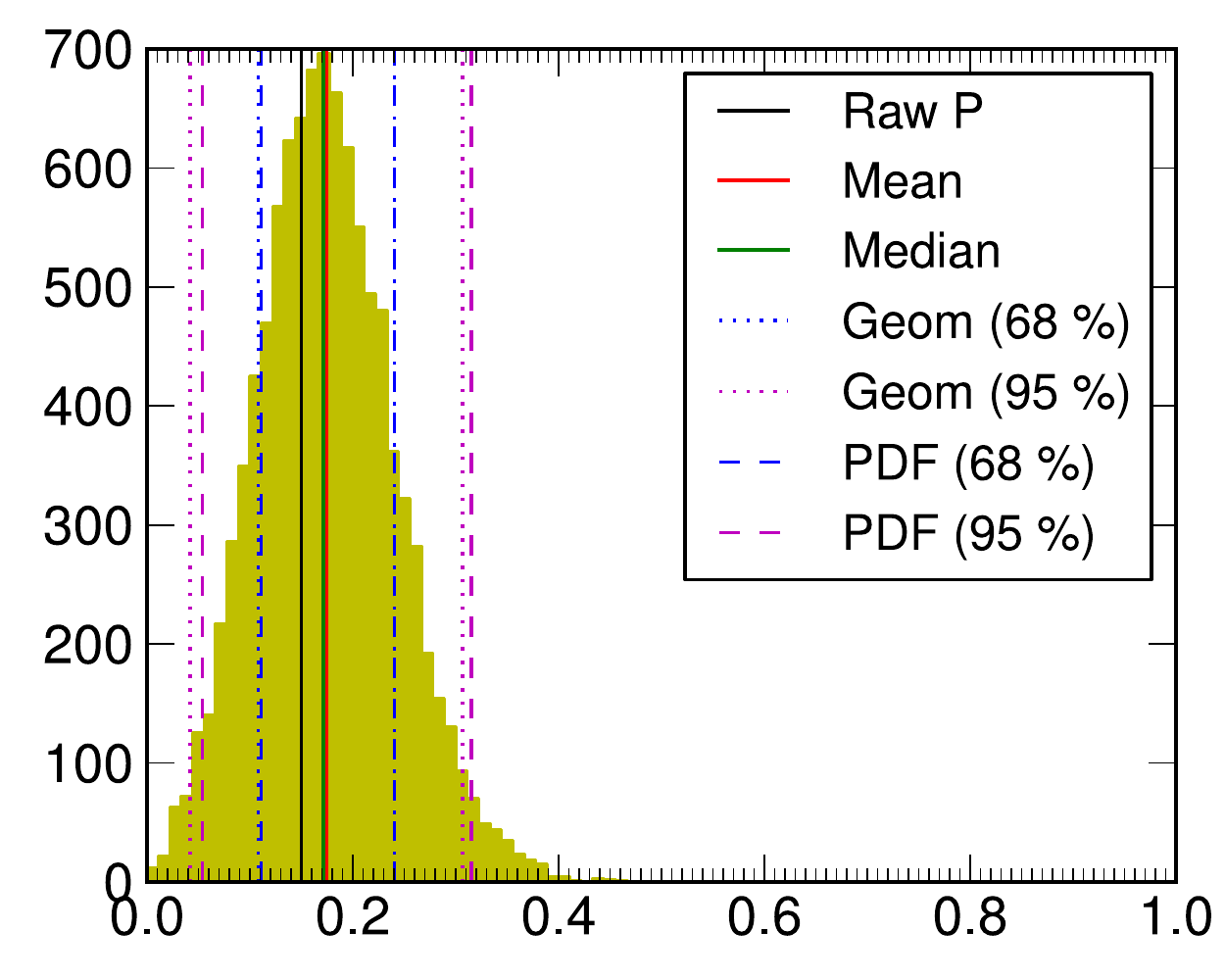}
\includegraphics[angle=0,width=54mm]{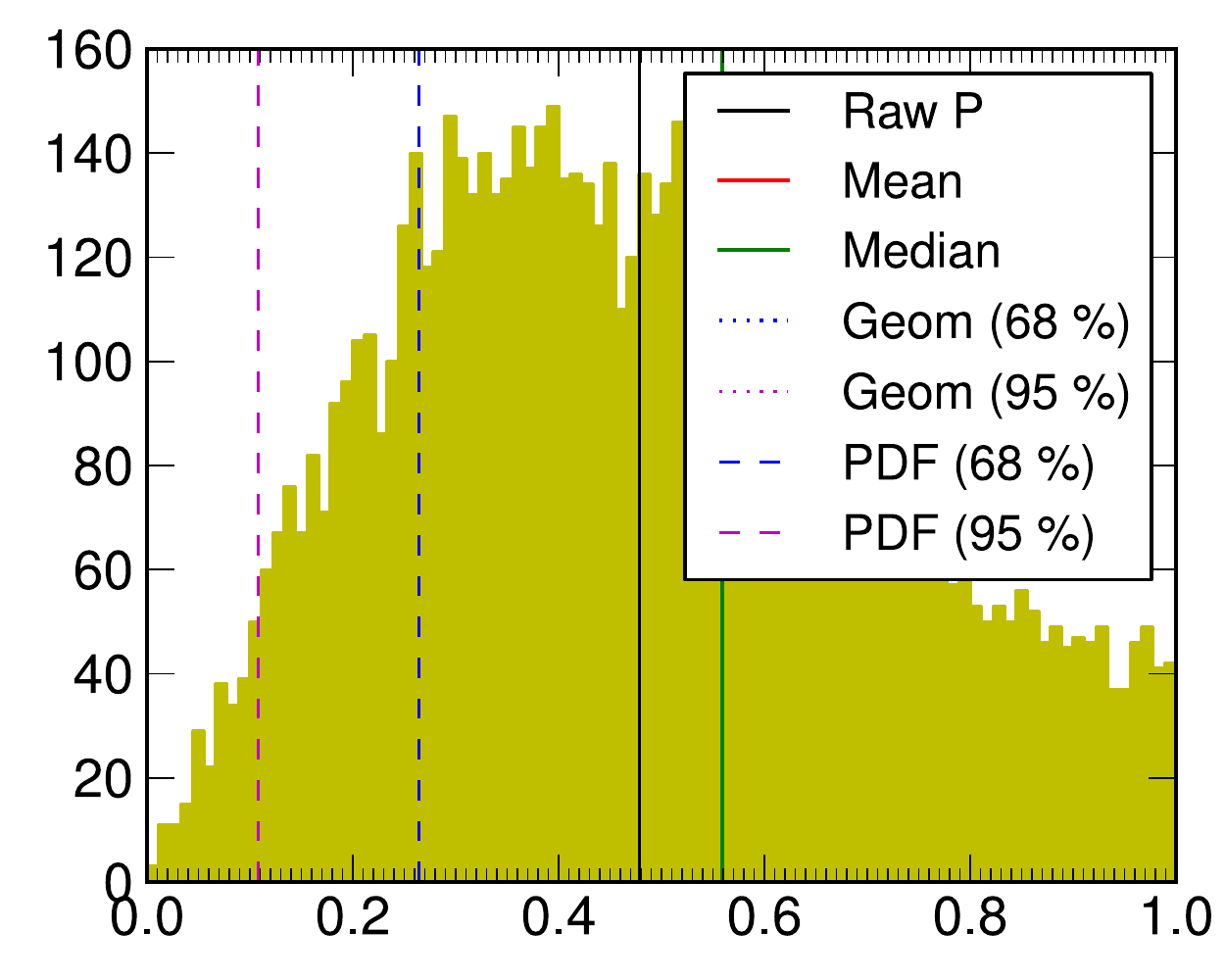}
\caption{~$\mathbf{|}$ \sffamily \textbf{The error distribution of polarization fraction $P$}.	
10,000 realizations of $P$ are determined by Monte Carlo
simulations at three different positions. Here we show the histogram of the 
realizations in yellow. The measured value of $P$, without Monte Carlo, is show
by black solid vertical lines, while the mean and median values of $P$ are shown in 
red and green respectively. In blue and magenta we show the confidence 
estimates made by taking the geometric limits (mean$\pm$ std) and by
integrating over the PDFs, as dotted and dashed lines respectively.
We show $1 \sigma$ limits in blue and $2\sigma$ in magenta. 
The {\em left} panel shows the central bin (the point of highest \lya\ 
surface brightness). Here the skewness of the distribution can just about be seen, 
		but the
polarization fraction is clearly well-constrained. $P$ is effectively consistent
with zero. 
The {\em central} panel shows a bin taken from the ring-like structure at 
a radial distance of 6 arcseconds: the distribution is largely symmetric
although skewness can still be seen. $P\approx 15$~per cent is measured, at
confidence above 95 per cent. The {\em right} panel shows a bin from the sky.
The distribution is skewed, mostly off the plot, and the geometrically determined 
confidence limits, even at 68 per cent, are outside the bounds of the
abscissa, demonstrating that no estimates of $P$ can be made.  }
\label{fig:disthist}
\end{figure*}
We first examine the observations of standard stars. We perform aperture 
photometry in each of the eight frames using the {\tt phot} task in 
{\tt IRAF}. We adopt homogeneous apertures of radius 1.5 arcseconds in each frame and
no re-centering. Further we obtain fluxes with and without subtraction of the 
local background. The calculations all derive from normalized differences 
(Equation~\ref{eq:normdiff}), and as described previously are not dependent 
upon
the absolute calibration of the data. However, imperfections in the Wollaston 
prism may cause the fraction of light not to be
split in exactly 50\% between the ordinary and extra-ordinary beams. We adopt
the magnitude and associated error output by {\tt phot} and compute $P$ and 
$\chi$ for both the polarized and unpolarized standards, with errors computed
as described above.

For the polarized standard we obtain $P=(4.33 \pm 0.14)$\% which is consistent 
with the quoted value of $P=4.32\pm 0.05$\%. 
For the unpolarized standard we obtain $P=(0.503 \pm 0.533)$\%. 
The errors reported by {\tt phot} are purely statistical 
and contain no systematic component but even so, the error estimates on $P$ 
are still consistent with the quotations without inflation of the error 
budget to account for a systematic component. This indicates that much of a 
systematic component, should it be present, is canceled in 
Equation~\ref{eq:normdiff}, which is very encouraging for our subsequent 
conclusions.
In summary we see no evidence for our measurements being subject to unequal 
splitting of light in the Wollaston prism and perform no correction for this. 

We next examine the sky background of the images to test for a spurious 
polarization signal that may result from improper flat-fielding or sky 
subtraction.  We define twenty square boxes of sides between 4 and 8 pixels, 
corresponding to $16-64$ pixels in total or $1-4$ square arcseconds. These 
sizes represent the approximate range of sizes covered by the binned resolution
elements, and do not exceed the maximum size set for the Voronoi tessellation.
Sky boxes are free from any astronomical sources in the deepest stacked science 
frame (all 88 raw frames). For each box we compute the mean and standard 
deviation of the sky in all eight input frames, and use these as input to compute
$P$ and $\chi$, using the expressions and Monte Carlo methods described 
previously. Firstly we find no preferred direction for
the polarization vector $\chi$, with a $\chi$ distribution that spans the 
range $0$ to $\pi$ radians. 
In our sky boxes we find a spurious polarization signal
at what would be deemed $1\sigma$ confidence in four boxes, which given the 
small numbers is roughly consistent with expectations. No boxes show
spurious detections of polarization at the $2\sigma$ confidence level, whereas 
many resolution elements surrounding LAB1 do. 

Naturally when we examine the science data we compute the Stokes parameters 
in individual bins using identical methods as for standard stars
and sky boxes described in the previous paragraphs. These results are reported in
the main article. In Figure~\ref{fig:disthist} of this supplementary information 
document we show the distribution 
of $P$ found by Monte Carlo in some individual bins of the science data, that 
illustrate a few typical cases: the central regions where the signal is very high
but the the \lya\ radiation is unpolarized (or polarized to a very low degree), 
the ring of strong polarization at a radius of $\approx 5$~arcseconds where the 
total signal is weaker but the polarized fraction higher, and the sky background
in the outer regions where there is effectively no \lya\ signal, resulting in a 
spuriously high raw measurement of $P$  but at no significance.

\section{Additional comments upon the results}

\subsection{The angle of polarization}

\begin{figure*}
\includegraphics[angle=0,width=120mm]{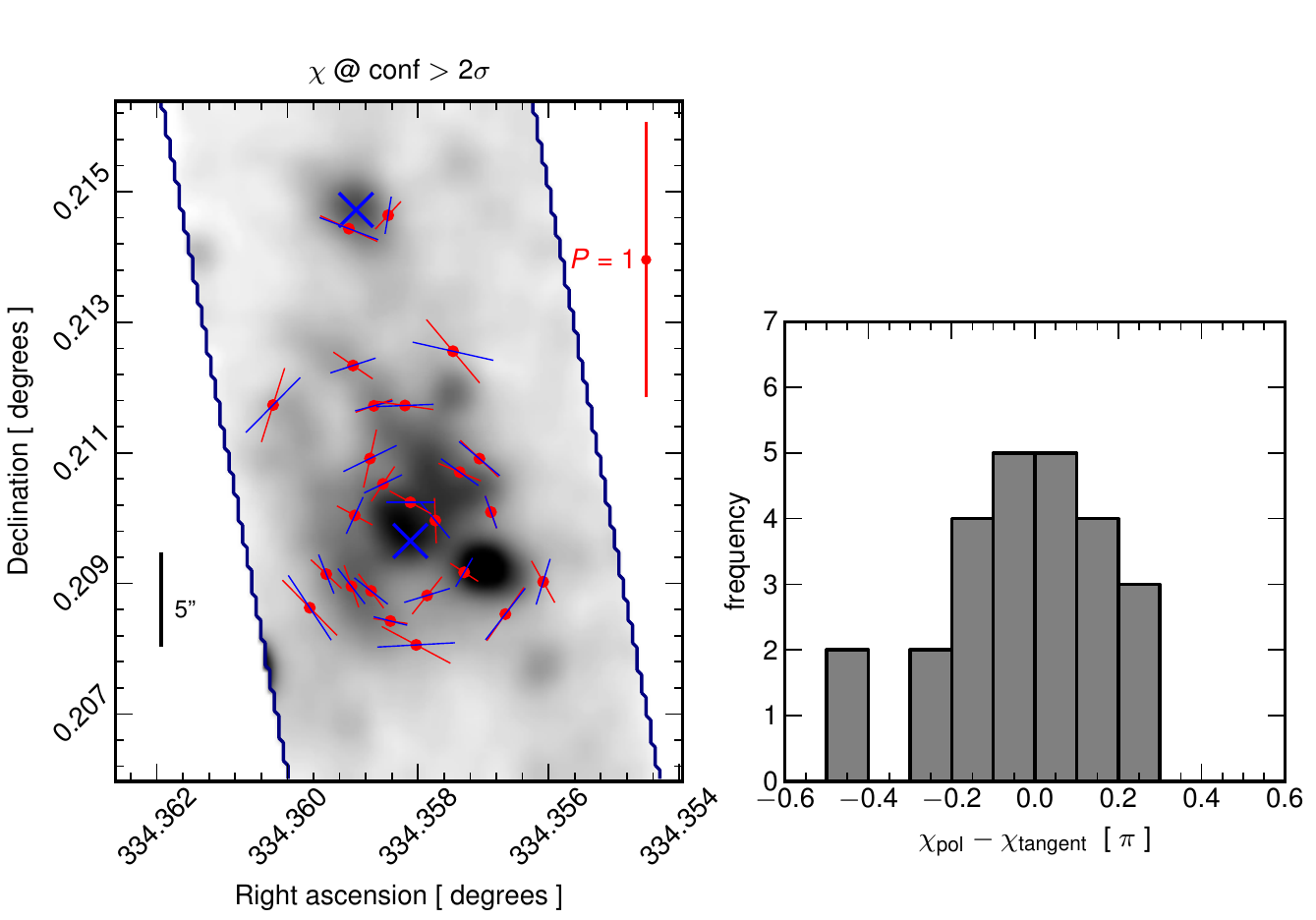}
\includegraphics[angle=0,width=120mm]{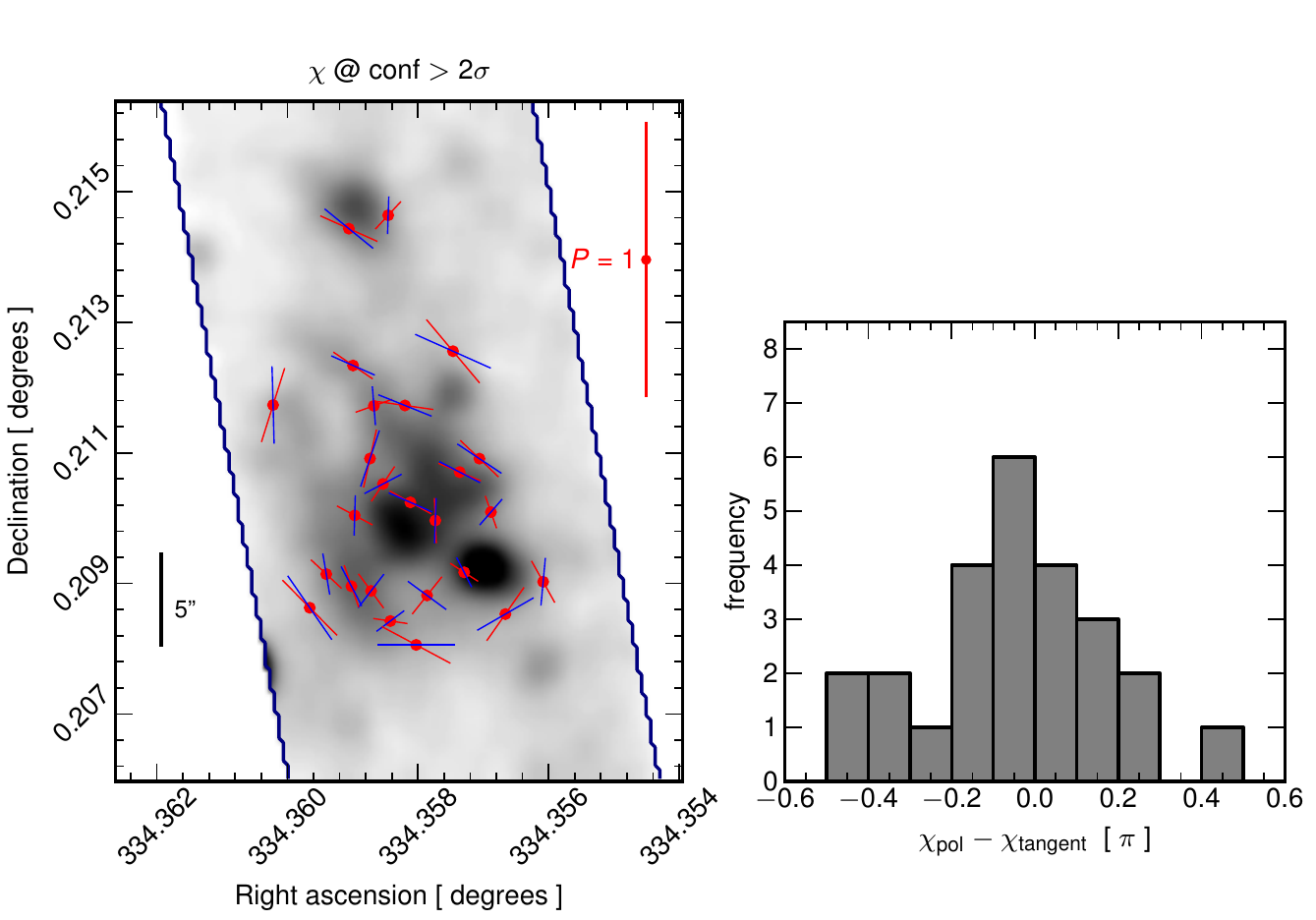}
\caption{~$\mathbf{|}$ \sffamily \textbf{The orientation of the angle $\chi$}. Both images show the 
smoothed image of the LAB with $\chi$ overlaid in red, as in Figure 2 of the 
main article. For each point we now over-plot in blue the tangential angle 
defined by the marked centres of LAB1 and LAB8 (upper) and the angle 
tangential to the local surface brightness contour (lower). The histogram 
to the side of each image shows the distribution of the difference between 
these two angles.  }
\label{fig:chidist}
\end{figure*}

In Figure~\ref{fig:chidist} we provide some more relevant information pertaining
to the polarization angle. It has been discussed in the main article that in
the case of spherical symmetry the polarization angle is expected to align
tangentially$^{17}$
to the overall structure. Of course in 
reality we do not expect spherical symmetry, which is well reflected by
the observed \lya\ surface brightness profile. Thus we may expect some hybrid 
situation in which $\chi$ is orientated around angles tangential to the local
surface brightness. 

In the left panel of Figure~\ref{fig:chidist} we display $\chi$ 
for each bin where the radiation is strongly polarized, and also the angle of 
the circular tangent to the overall system. The centre is defined as the 
brightest pixel in the narrowband image that is not associated with the AGN. 
By eye it is immediately visible that the $\chi$ is aligning tangentially with
the overall geometry of the system (for this experiment we also define a centre
for LAB8 to the north of the image). We show also a histogram of the difference
between these angles; if the polarization vector were randomly orientated we 
would expect a flat distribution of angle differences that ranges between 
$-\pi/2$ and $-\pi/2$. Certainly this is not what what the histogram shows, and
$\chi$ minus the tangent angle is strongly clustered around zero.

The right panel of Figure~\ref{fig:chidist} is the same, except instead of the
tangent to circles centred at the origin, we display the angle tangential to the
local surface brightness contour. The situation is effectively the same as in 
the plot to the left, and clearly the polarization vectors are aligned far from 
randomly, with a peak at zero.

\newpage

\section{Consistency with recent measurements}

Numerous studies have addressed the polarized emission from high-redshift 
AGN, which are frequently found to be polarized in the continuum but not 
\lya\cite{Cimatti1998}. 
Only one other study of the \lya\ polarization signal from a LAB 
exists (Prescott et al, 2011)$^{23}$, 
targeting LABd05\cite{Dey2005}, 
which does host an AGN within the halo.
 This study also
adopted imaging polarimetry and an observational method and strategy that 
was very similar to our own: a narrowband filter; four positions of a half wave
plate retarder; and 16 hours of integration (compared with 14.5 for our 
programme).  This study found the \lya\ from LABd05 to be unpolarized. 

However there are a number of substantial differences between the pair of 
observations: 
they used the 2.3m Bok telescope while we used the 8.2m VLT; 
the pixel scale of Bok/SPOL is 0.5 arcsec/px, while that of VLT/FORS is 0.25 arcsec/px;
their seeing was typically 1.5--2.3 arcsec while our average is 0.96 arcsec; 
and their individual
integrations were 200 seconds while ours were 1,200. In the regime where the noise is
dominated by the sky background, $S/N$ scales linearly with the diameter of the primary mirror
and the number of pixels contained within a seeing disk. Given that the pixel scales of 
FORS and SPOL differ by a factor of 2, and the respective seeings differ by the same value, 
we expect our $S/N$ to be better by a factor of the ratio of the mirror diameters,
$8.2/2.3\approx 4$. 
However, this factor is likely to be a conservative estimate because the
short exposures employed by Prescott et al would most likely result in noise 
that is dominated by CCD read-out, not sky. 

As previously stated, the signal-to-noise on $P$ goes as 
$(S/N)_P = \sqrt{N_\mathrm{HWP}/2} \cdot P \cdot (S/N)_\theta$, which would imply that
were we to observe the same target with the configuration of Prescott et al, we 
would have obtained a similarly tight non-detection over the inner few arcsec, but would
not have been able to probe to the radii where polarized radiation is seen.

\clearpage